\documentclass[aps,prd,twocolumn,floatfix,superscriptaddress,preprintnumbers,nofootinbib]{revtex4-1}

\input{header}
\usepackage{ulem}
\usepackage{float}
\usepackage{lipsum}

\newcommand{\hD}{h_D}
\def\triumf{TRIUMF, 4004 Wesbrook Mall, Vancouver, BC V6T 2A3, Canada}
\def\sfu{Department of Physics, Simon Fraser University, Burnaby, BC V5A 1S6, Canada}

\begin{document}
\date{\today}
\title{Higgs Portal Interpretation of the Belle II $B^+ \to K^+ \nu \nu$ Measurement} 

\author{David McKeen}
\email{mckeen@triumf.ca}
\affiliation{\triumf}
\author{John N. Ng}
\email{misery@triumf.ca}
\affiliation{\triumf}
\author{Douglas Tuckler}
\email{dtuckler@triumf.ca}
\affiliation{\triumf}
\affiliation{\sfu}

\begin{abstract}
The Belle II experiment recently observed the decay $B^+ \to K^+ \nu \nu$ for the first time, with a measured value for the branching ratio of $ (2.3 \pm 0.7) \times 10^{-5}$. This result exhibits a $\sim 3\sigma$ deviation from the Standard Model (SM) prediction. The observed enhancement with respect to the Standard Model could indicate the presence of invisible light new physics. In this paper, we investigate whether this result can be accommodated in a minimal Higgs portal model, where the SM is extended by a singlet Higgs scalar that decays invisibly to dark sector states. We find that current and future bounds on invisible decays of the 125 GeV Higgs boson completely exclude a new scalar with a mass $\gtrsim 10$ GeV. On the other hand, the Belle II results can be successfully accommodated if the new scalar is lighter than $B$ mesons but heavier than kaons. We also investigate the cosmological implications of the new states and explore the possibility that they are part of an abelian Higgs extension of the SM. Future Higgs factories are expected to place stringent bounds on the invisible branching ratio of the 125 GeV Higgs boson, and will be able to definitively test the region of parameter space favored by the Belle II results. 
\end{abstract}

\maketitle

\section{Introduction}
Semi-leptonic decays of $B$ mesons mediated by flavor changing neutral currents (FCNCs) are extremely rare in the Standard Model (SM), and are some of the cleanest probes of beyond-the-SM physics (BSM). FCNC decays of $B$ mesons to charged leptons, such as $b \to s \ell\ell$ transitions, are in good agreement with SM prediction and have placed stringent constraints on BSM particles coupled to leptons. On the other hand, FCNC decays involving neutrinos in the final state are experimentally challenging to measure because of the missing energy carried away by SM neutrinos. Bounds from, for example, Belle on these semi-invisible decays have been studied in the past and led to stringent constraints on various BSM models.

Recently, the Belle II collaboration reported the first evidence of the rare decay $B^+ \to K^+ \nu \bar{\nu}$. The branching ratio (BR) was measured with two methods: a conventional hadronic-tag method and an novel inclusive-tag method. The measured branching ratios using these two methods are found to be \cite{BelleII:EPS1,BelleII:EPS2,Belle-II:2023esi}
\begin{align}
\text{BR}(B^+ \to K^+ \nu \bar{\nu})_\text{had} &= ( 1.1^{+0.9}_{-0.8} {}^{+0.8}_{-0.5}) \times 10^{-5},\\
\text{BR}(B^+ \to K^+ \nu \bar{\nu})_\text{incl} &=  (2.7\pm 0.5 \pm 0.5) \times 10^{-5},
\end{align}
where the first and second uncertainties are statistical and systematic uncertainties, respectively.
A combination of these two measurements yields the final result
\begin{equation}
\text{BR}(B^+ \to K^+ \nu \bar{\nu})_\text{exp} = (2.3 \pm 0.7) \times 10^{-5}.
\end{equation}

The Standard Model prediction for this rare decay is \cite{Bause:2023mfe}
\begin{equation}
\text{BR}(B^+ \to K^+ \nu \bar{\nu})_\text{SM} = (4.29 \pm 0.23) \times 10^{-6},
\end{equation}
where the the tree-level contribution from $B^+\to \tau^+(\to K^+\bar{\nu})\nu$ has been subtracted. We can immediately see that the Belle II measurement using the hadronic tag is consistent with the SM, while the inclusive measurement has a $3.6\sigma$ tension with the SM prediction. The combined result is in tension with the SM at 2.8$\sigma$.

This result can be interpreted as the presence of new particles produced in $B^+ \to K^+$ decays that are invisible on the length scale of the Belle II detector, such as new physics coupled to neutrinos or dark matter (see \cite{Athron:2023hmz,Bause:2023mfe,Allwicher:2023syp,He:2023bnk,Felkl:2023ayn,Datta:2023iln,He:2023bnk,Berezhnoy:2023rxx} for recent studies). In particular, a new \textit{light} particle $X$ that couples to a dark sector state $\chi$ can be produced in the two-body decay $B^+ \to K^+ X$  if $m_X \lesssim m_B$, with $X$ decaying invisibly to dark sector states. On the other hand, $X$ can mediate the three-body decay $B^+ \to K^+ \chi \bar{\chi}$ if $m_X > m_B$. The presence of these light new particles can enhance  $B^+ \to K^+  + {\rm inv.}$ and provide an explanation for the Belle II result.

The enhancement compared to the SM prediction is found by taking the difference between the Belle II combined result and the SM prediction. We find that the BR for the new physics contribution is
\begin{equation}\label{eq:BelleNP}
\text{BR}(B^+ \to K^+ + {\rm inv.})_\text{NP} = (1.9 \pm 0.7) \times 10^{-5}.
\end{equation}

In this paper, we consider the possibility that a singlet scalar $S$ could accommodate the Belle II measurement. In this scenario, a new scalar $S$ mixes with the SM Higgs $h$ and communicates to the dark sector via renormalizable interactions. The $h-S$ mixing induces couplings of $S$ to SM fermions and new contributions to $B^+ \to K^+  + {\rm inv.}$ will arise when $S$ decays invisibly to dark sector states. We consider two scenarios for the dark sector:
\begin{itemize}
\item[1.]\textit{Singlet Higgs Portal:} the dark sector consists of the scalar $S$ and dark fermions $\chi$, and are both are pure singlets (i.e. no additional dark gauge group). We find that $\chi$ particles come into thermal equilibrium with the SM bath and have a relic abundance that is larger than the observed dark matter abundance. Hence, the dark fermions have to either be unstable or need additional annihilation channels. 

\item[2.] \textit{Dark Abelian Higgs Model:} the dark sector states are dark photons $A_D$  and a dark Higgs $\Phi_D$ that arise in a $U(1)_D$ extension of the SM. In this scenario, the singlet Higgs carries a $U(1)_D$ charge and a vacuum expectation value (VEV) that spontaneously breaks the dark gauge symmetry. In addition, the dark photon kinetically mixes with the SM photon and is naturally unstable. For a sufficiently long lifetime, the dark photon appears as missing energy at Belle II. 
\end{itemize}

In both scenarios, we find that if $m_S \gtrsim 5$ GeV, the Belle II measurement can be accommodated when the mixing angle is relatively large $\mathcal{O}(0.1-1)$. However, this is in tension with current constraints on the invisible branching ratio of the SM Higgs and is completely ruled out. This could be alleviated in more extended Higgs sector models (e.g. two Higgs double models), where the singlet mixes dominantly with the additional Higgs bosons rather than the SM-like Higgs \cite{Ipek:2014gua,Batell:2016ove}. In this way, invisible SM Higgs decays can be avoided.

On the other hand, a light scalar with $m_S \lesssim m_B$ can explain the Belle II results while being consistent with current bounds on invisible SM Higgs decays. However, stringent bounds on invisible kaon decays rule out the parameter space with $m_S \lesssim m_K$. In the region $m_K \lesssim m_S \lesssim m_B$, we find that the Belle II result can be successfully explained while being consistent with bounds on other $B$ meson decays such as $B^0 \to K^{\ast 0} \nu \nu$. 

In the future, the high-luminosity run of the LHC (HL-LHC) is expected to constrain the SM Higgs invisible BR to be $< 0.025$ and will rule out some of parameter space that can explain the Belle II result. Future Higgs factories such at the International Linear Collider (ILC) \cite{ILCInternationalDevelopmentTeam:2022izu}, the Compact Linear Collider (CLIC) \cite{Brunner:2022usy}, or the Future Circular Collider (FCCee) \cite{FCC:2018evy} are projected to constrain the invisible branching ratio of the SM HIggs to the sub-percent level, and can definitively test the parameter space of the Higgs portal model. 

This paper is organized as follows. In Sec.~\ref{sec:HiggsPortal}, we discuss the singlet Higgs portal model, the contribution to $B^+ \to K^+  + {\rm inv.}$, and additional constraints arising from SM Higgs invisible decays and other meson decays. In Sec.~\ref{sec:DM} we briefly discuss the early universe cosmology of the dark fermions of the singlet Higgs portal model. Sec.~\ref{sec:DarkHiggs} is devoted to Dark Abelian Higgs model. We conclude in Sec.~\ref{sec:conclusion}.

\section{Singlet Higgs Portal Model}\label{sec:HiggsPortal}

We begin by introducing the singlet Higgs portal model. We extend the Standard Model (SM) by an additional singlet scalar field $S$ that mixes with the SM Higgs boson $H$ via the renormalizable interaction \cite{Krnjaic:2015mbs}
\begin{equation}\label{eq:Lag}
\mathcal{L} \supset -A_{S H} S H^\dagger H
\end{equation}
and we assume that $S$ does not get a vacuum expectation value.\footnote{We will discuss the possibility that the scalar develops a vacuum expectation value as well as the impact of the operator $S^2H^\dagger H$ in Sec.~\ref{sec:DarkHiggs}.}

In addition, we add a dark sector Dirac fermion $\chi$ with mass $m_\chi$ that interacts with the new scalar via

\begin{equation}
\mathcal{L}_D = y_D S \bar{\chi} \chi.
\end{equation}

After electroweak symmetry breaking, $S$ and $H$ mix to form mass eigenstates $h$ and $h_D$ given by

\begin{align}\label{eq:mixing}
H &= h\cos\theta  + h_D\sin\theta,  \\
S &=  -h \sin\theta  + h_D\cos\theta 
\end{align}

where the mixing angle $\theta$ diagonalizes the scalar mass-squared matrix. We identify $h$ as the SM Higgs boson discovered at the LHC with a mass $m_h = 125$ GeV, and the dark Higgs $h_D$ as the mediator between the dark sector and the SM. In terms of the physical scalars, the couplings to SM fermions and dark fermions are
\begin{equation}
\begin{aligned}\label{eq:Yuk}
\mathcal{L} \supset h\cos\theta  &\sum_f  \frac{m_f}{v}\bar{f}{f} - h \sin\theta y_D\bar{\chi}\chi\\
 + h_D\sin\theta  &\sum_f  \frac{m_f}{v}\bar{f}{f} + h_D \cos\theta y_D\bar{\chi}\chi 
\end{aligned}
\end{equation}
where $m_f$ is the mass of the SM fermion $f$ and $v = 246$ GeV is the SM Higgs vacuum expectation value (VEV).

The interactions in Eq.~\eqref{eq:Yuk} lead to FCNC decays of kaons and $B$ mesons mainly via a top-quark loop. The effective Lagrangian for FCNC couplings of the Higgs bosons to down quarks $d_{i,j}$ is

\begin{equation}\label{eq:Leff}
\mathcal{L}_\text{eff} = c_{ij} ( h_D\sin\theta + h \cos\theta ) \bar{d_j} P_R d_i + \text{h.c.},
\end{equation}
 
 where 
 
 \begin{equation}
 c_{ij} = \frac{3\sqrt{2}}{16 \pi^2} \frac{G_F m_i m^2_t}{v} V_{ti} V^\ast_{tj} \end{equation}
is the effective coupling after integrating out the top quark and $W^\pm$ boson in the loop.  $V_{ti,tj}$ are CKM matrix elements, $m_t$ is the top quark mass, and $m_i$ is the mass of the initial state quark.


\subsection{Singlet Higgs Contribution to $B^+ \to K^+ + {\rm inv.}$ }\label{sec:Belle}

\begin{figure*}[tbh]
\includegraphics[width = 0.55\textwidth]{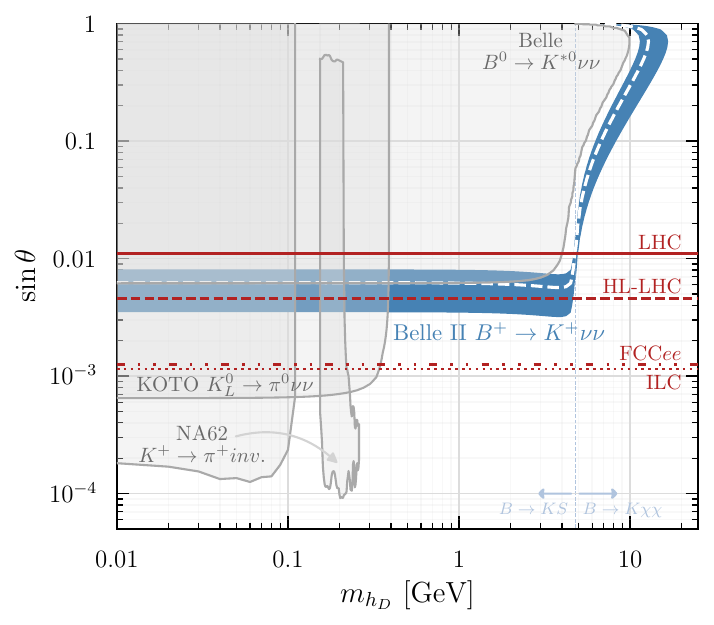}
\caption{Region of $\sin\theta$ vs $m_S$ plane that can explain the Belle II excess at 2$\sigma$ (blue shaded region) for the singlet Higgs portal model. The central value from Eq.~\eqref{eq:BelleNP} is depicted by the dashed white curve, and  the red horizontal lines are current and future bounds on the invisible BR of the SM Higgs boson in Tab.~\ref{tab:Hinv}.  We show additional constraints from $K^+ \to \pi^+ + {\rm inv.}$ \cite{NA62:2021zjw}, $K_L^0 \to \pi^0 \nu\nu$ \cite{KOTO:2020prk} , and $B^0 \to K^{\ast 0} \nu\nu$ \cite{Belle:2017oht} in the gray shaded regions. The vertical dotted blue line denotes the transition from two-body decays $B \to K S$ to three-body decays $B \to K \chi\chi$.} \label{fig:bounds}
\end{figure*}

Together, the interactions in Eqs.~\eqref{eq:Yuk} and \eqref{eq:Leff} contribute to decays of $B$ mesons and kaons involving missing energy when $y_D \gtrsim (m_\mu/v)\sin\theta$. In this regime, the dark Higgs $h_D$ decays dominantly to a pair of dark fermions. When $m_{\hD} < (m_B- m_K)$, $B$ mesons undergo two-body decays to kaons and on-shell dark Higgs bosons, while three-body decays are mediated by off-shell Higgs bosons when $m_{\hD} > m_B$. Similarly, the dark Higgs is produced on-shell in kaon decays when $m_{\hD} < (m_K- m_\pi)$. The rates for these two-body decays are
\begin{widetext}
\begin{align}
\Gamma(B^+ \to K^+ \hD) &= \frac{|c_{bs}|^2\sin^2\theta}{64 \pi m_{B^+}^3} |f^{BK}_0(m_{\hD}^2)|^2\bigg(\frac{m_{B^+}^2 - m_{K^+}^2}{m_b - m_s}\bigg)^2\lambda^{1/2}(m_{B^+}^2,m_{K^+}^2,m_{\hD}^2), \label{eq:BpltoKpl}\\
\Gamma(B^0 \to K^{\ast 0} \hD) &=  \frac{|c_{bs}|^2\sin^2\theta}{64 \pi m_{B^0}^3} |A^{BK}_0(m_{\hD}^2)|^2\bigg(\frac{1}{m_b + m_s}\bigg)^2\lambda^{3/2}(m_{B^0}^2,m_{K^{\ast 0}}^2,m_{\hD}^2), \label{eq:B0toK0}\\
\Gamma(K\to \pi \hD) &=  \frac{|c_{sd}|^2\sin^2\theta}{64 \pi m_K^3} |f^{K\pi}_0(m_{\hD}^2)|^2\bigg(\frac{m_K^2 - m_\pi^2}{m_s - m_d}\bigg)^2\lambda^{1/2}(m_K^2,m_\pi^2,m_{\hD}^2) \label{eq:KtoPi},
\end{align}\end{widetext}
where $f_0(q^2)$ and $A_0(q^2)$ are hadronic form factors for $B \to K$ or $K \to \pi$ transitions that are evaluated at the mass of the dark Higgs $m_{\hD}$.  The parameterization of these form factors is discussed in App.~\ref{app:formfactors}. The function $\lambda(a,b,c) = a^2 + b^2 + c^2 - 2(ab + bc + ac)$ is the K\"{a}ll\'{e}n function. In Eq.~\eqref{eq:KtoPi}, $K \in \{K^+, K^0\}$ and $\pi \in \{\pi^+, \pi^0\}$.

The differential rates for the three-body decays $B^+\to K^+ \chi \bar{\chi}$ and $B^0 \to K^{\ast 0}\chi \bar{\chi}$ are

\begin{widetext}
\begin{align}
\frac{d\Gamma}{dq^2} (B^+ \to K^+ \chi \bar{\chi})&= \frac{q^2 |f^{BK}_0(q^2)|^2 }{512 \pi^3 m_{B^+}^3} 
\bigg( \frac{m_{B^+}^2 - m_{K^+}^2}{m_b - m_s}\bigg)^2  
\bigg|\frac{c_{bs}\sin\theta\cos\theta}{q^2 - m^2_{\hD} + i m_{\hD} \Gamma_{\hD}} - \frac{c_{bs}\sin\theta\cos\theta}{q^2 - m^2_h}\bigg|^2 \nonumber\\
&\hspace{6cm}\times \bigg(1 - \frac{4m_\chi^2}{q^2}\bigg)^{3/2}\lambda^{1/2}(q^2,m_{B^+}^2, m_{K^+}^2),\\
\frac{d\Gamma}{dq^2} (B^0 \to K^{\ast 0} \chi \bar{\chi})&= \frac{q^2 |A^{BK}_0(q^2)|^2}{512 \pi^3 m_{B^0}^3}\frac{1}{(m_b + m_s)^2}
\bigg|\frac{c_{bs}\sin\theta\cos\theta}{q^2 - m^2_{\hD} + i m_{\hD} \Gamma_{\hD}} - \frac{c_{bs}\sin\theta\cos\theta}{q^2 - m^2_h}\bigg|^2\nonumber\\
&\hspace{6cm}\times  \bigg(1 - \frac{4m_\chi^2}{q^2}\bigg)^{3/2} \lambda^{3/2}(q^2,m_{B^0}^2, m_{K^{\ast 0}}^2).
\end{align}
\end{widetext}

The total rate for three-body decays is found by integrating these from $q^2_\text{min} = 4m_{\chi}^2$ to $q^2_\text{max} = (m_B - m_K)^2$. In the equations above, $\Gamma_{\hD}$ is the total width of the $\hD$ which is dominated by the decay rate into the dark fermions and is

\begin{equation}\label{eq:Sinv}
\Gamma(\hD \to \chi \bar{\chi}) = \frac{ y^2_D \cos^2\theta}{8\pi} m_{\hD} \bigg(1- \frac{4 m_\chi^2}{m_{\hD}^2}\bigg)^{3/2}.
\end{equation}

\begin{figure}[tbh]
\includegraphics[width = 0.45\textwidth]{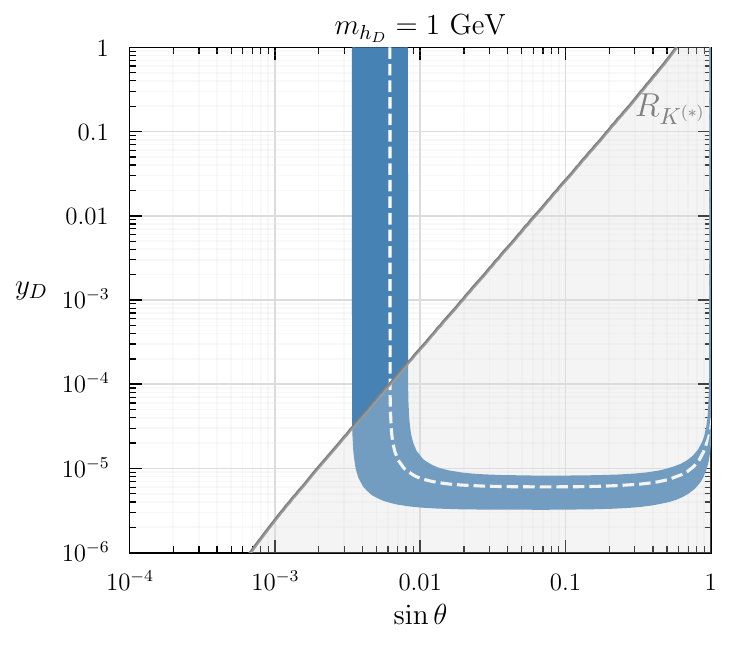}
\caption{Region of $y_D$ vs $\sin\theta$ plane that can explain the Belle II result at 2$\sigma$ (blue shaded region) in the singlet Higgs portal with a dark Higgs mass $m_{\hD} = 1$ GeV. The central value from Belle II is depicted by the dashed white curve. Constraints from LFU tests $R_{K^{(\ast)}}$ are depicted by the gray shaded regions.} \label{fig:yD}
\end{figure}

There are decay modes to SM fermions that are proportional to $(m_f/v) \sin\theta$ and are subdominant when $y_D >(m_f/v)\sin\theta$. This can be seen in Fig.~\ref{fig:yD} where we show the region of $y_D -\sin\theta$ that can explain the Belle II excess in the blue shaded region. As $y_D$ gets smaller, the decay to  muons starts to become important. Because the couplings of the dark Higgs are proportional to the fermion masses, there is an inherent lepton flavor non-universality which is severely constrained by LHCb measurements of $R_K$ and $R_{K^\ast}$, which are in agreement with SM predictions \cite{LHCb:2022qnv,LHCb:2022vje}. The bounds on the $y_D$ vs $\sin\theta$ parameter from  tests in $b\to s$ transitions are depicted by the gray shaded region in Fig.~\ref{fig:yD} and exclude $y_D \lesssim 10^{-4}$. 

The main result of this section is in Fig.~\ref{fig:bounds}, where we show the region of $\sin\theta - m_{\hD}$ plane that can fit the Belle II excess at the $2\sigma$ level (blue shaded region). We see that a dark Higgs with $m_{\hD} \lesssim m_B$ can explain the Belle II excess with a mixing angle of $\sin\theta \simeq 6 \times 10^{-3}$, while a heavier dark Higgs requires larger mixing angles.

\subsection{Higgs Invisible Decays}\label{sec:higgsinv}
After mixing, the SM Higgs couples to the dark sector particles with strength $y_D \sin\theta$. For $m_\chi < m_h/2$, this leads to invisible decays of the SM Higgs boson. The decay width for this process is
\begin{equation}\label{eq:HiggsInv}
\Gamma(h\to \chi \chi) = \frac{y^2_D \sin^2\theta}{8\pi} m_h \bigg(1 - \frac{4m_\chi^2}{m_h^2}\bigg)^{3/2}.
\end{equation}

Currently, the strongest bound on invisible  decays of the SM Higgs boson are from LHC with BR$^h_{\rm inv}<$  0.13 \cite{ATLAS:2023tkt,CMS:2023sdw}. In the future, HL-LHC is expected to constrain the invisible BR$^h_\text{inv}$ to be less than 2.5\% \cite{Cepeda:2019klc}. Using the expression in Eq.~\eqref{eq:HiggsInv} and the SM prediction for the total width of the Higgs $\Gamma_h = 4.1$ MeV, we can calculate the BR for $h\to \chi\chi$ as

\begin{equation}
\text{BR}(h \to \chi \chi) = \frac{\Gamma(h\to \chi \chi)}{\Gamma_h + \Gamma(h\to \chi \chi)}.
\end{equation}

It is straightforward to calculate a bound the SM Higgs couplings to dark sector fermions. We find that the mixing angle must satisfy

\begin{equation}
\sin\theta \lesssim \sqrt{\frac{8\pi}{m_h} \text{BR}^h_\text{inv}\Gamma_h} = 0.01~(4.5\times10^{-3})
\end{equation}
for LHC (HL-LHC) bounds. The bounds on this scenario are depicted in Fig.~\ref{fig:bounds} by the red solid and dashed lines for the LHC and HL-LHC, respectively. Current LHC bounds rule out the region of parameter space the can address the Belle II result when $m_{\hD}\gtrsim m_B$, but are not strong enough to rule out smaller $m_{\hD}$ masses. HL-LHC will probe most of the favored parameter space.

The Higgs invisible BR is expected to be constrained to be less than a percent at future Higgs factories, like the ILC \cite{Potter:2022shg}, CLIC \cite{CLICdp:2018cto}, and FCC$ee$ \cite{Blondel:2021ema}. The strongest of these are from ILC and FCC$ee$ at 0.16\% and 0.19\%, respectively, and are depicted in Fig.~\ref{fig:bounds} by the dotted and dot-dashed red lines. 

If the Belle II result on $B^+ \to K^+ \nu \nu$ is confirmed in the future, then these experiments will be able to probe the favored parameter space for the light singlet Higgs scenario. The singlet Higgs portal model is an essential benchmark for future Higgs factories, as improved bounds on (or a measurement of) BR$^h_\text{inv}$ will conclusively rule out or confirm the model.

\begin{table}[tbh]
\begin{ruledtabular}
\begin{tabular}{l l}
LHC & 13\% \cite{ATLAS:2023tkt,CMS:2023sdw}\\
HL-LHC & 2.25\% \cite{Cepeda:2019klc}\\
ILC & 0.16\% \cite{Potter:2022shg}\\
CLIC & 0.69\% \cite{CLICdp:2018cto}\\
FCC$ee$ &  0.19\% \cite{Blondel:2021ema}\\
\end{tabular}
\end{ruledtabular}
\caption{\label{tab:Hinv} Current and future bounds on Higgs invisible BR from the LHC\cite{ATLAS:2023tkt,CMS:2023sdw}, HL-LHC  \cite{Cepeda:2019klc}, ILC \cite{Potter:2022shg}, CLIC \cite{CLICdp:2018cto}, and FCC$ee$  \cite{Blondel:2021ema}.}
\end{table}

\subsection{$B^0 \to K^{\ast 0}$ Decays}

The effective operator in Eq.~\eqref{eq:Leff} also generates $B$ decays to vector mesons. In particular, the decay $B^0 \to K^{\ast 0}\nu\nu$ has been searched for by Belle and they found an upper bound on the branching ratio of \cite{Belle:2017oht}
\begin{equation}
\text{BR}(B^0 \to K^{\ast 0} \nu \nu) < 1.8 \times 10^{-5}.
\end{equation}
For $m_{\hD} < (m_B - m_K)$, this bound yields an additional constraint on the minimal Higgs model
\begin{equation}
\sin\theta \lesssim 6.2 \times 10^{-3}.
\end{equation}
This bound is shown in Fig.~\ref{fig:bounds} by the gray shaded region region labeled ``Belle II $B^0 \to K^{\ast 0} \nu\nu$, and is beginning to probe the blue shaded region that can explain the Belle II result, excluding the central value for $m_{h_D} \lesssim 1$ GeV.

\subsection{Kaon Decays}\label{sec:kaons}
In addition to contributions to rare $B$ meson decays, the $h-S$ mixing generates contributions to rare kaon decays, such as $K^+ \to \pi^+ +{\rm inv.}$ and $K_L^0 \to \pi^0 + {\rm inv.}$ Such decays are stringently constrained by the NA62\cite{NA62:2021zjw} and KOTO experiments \cite{KOTO:2018dsc,KOTO:2020prk}, respectively. 

Recently, the NA62 experiment measured the BR for the decay $K^+ \to \pi^+ \nu\nu$ to be $(10.6 ^{+4.0}_{-3.4} \pm 0.9)\times 10^{-11}$ at 68$\%$ C.L. \cite{NA62:2021zjw}. This measurement was used to set limits on the decay $K^+ \to \pi^+ X$ where $X$ is a (pseudo) scalar boson. Upper limits on this decay are depicted in the left plot Fig.~\ref{fig:bounds} by the gray shaded region label ``NA62 $K^+ \to \pi^+ +{\rm inv.}$ We see that this excludes the favored parameter space for $m_{\hD} \lesssim$ 200 MeV.

In addition, this same interaction leads to $K_L^0 \to \pi^0 + {\rm inv.}$ decays, mimicking  $K_L^0 \to \pi^0 +\nu \nu$. The latter decay has recently searched for by the KOTO experiment and an upper limit on the BR was found to be
\begin{equation}
\text{BR}(K_L^0 \to \pi^0 \nu \nu) < 3.0\times 10^{-9}.
\end{equation}
We take this value as an upper limit on the $K_L^0 \to \pi^0 S$ ($K_L^0 \to \pi^0 \chi\chi$) decay for the light (heavy) Higgs case. Using Eq.~\eqref{eq:B0toK0}, we obtain the approximate upper bound for the lights scalar case of
\begin{equation}
\sin\theta \lesssim 6.5 \times 10^{-4}.
\end{equation}

This bound is depicted in Fig.~\ref{fig:bounds} by the gray shaded region labeled ``KOTO $K_L^0 \to \pi^0 \nu \nu$'', and excludes $m_{h_D} \lesssim 400$ MeV.

To summarize the results, the preferred region of parameter space that fits the Belle II excess is a dark Higgs with a mass $ m_K\lesssim m_{\hD} \lesssim m_B$ and a mixing angle with the SM Higgs of $\sin\theta \simeq 6\times 10^{-6}$.

\section{Can the dark fermions be the DM?} \label{sec:DM}

The dark fermions so far have no role beyond inducing invisible $\hD$ decays. A natural question to ask is: can the dark fermions constitute all of the DM of the universe? To answer this question, we estimate the present-day $\chi$ abundance in this simple model.

\subsection{Freeze-out Abundance}

As a first step we determine the value of the mixing angle sufficient to bring the singlet scalar into equilibrium with the SM plasma through inverse decays by equating the rate for $h_D\to f\bar f$ to the value of Hubble expansion rate at $T=m_S$,
\begin{equation}
H(T) = 1.67\sqrt{g_\ast}\frac{T^2}{M_\text{Pl}} \simeq  10^{-18}\bigg(\frac{T}{\text{GeV}}\bigg)^2.
\end{equation}
where $g_\ast(T)$ is the effective number of relativistic degrees of freedom at $T$ and $M_{\rm Pl}\simeq1.2\times10^{19}~\rm GeV$ is the Planck mass. Doing so we find a critical value for the mixing angle of
\begin{equation}
\sin\theta_\text{cr} \sim 10^{-5} \bigg(\frac{m_\mu}{m_f}\bigg)\bigg(\frac{m_{\hD}}{\text{GeV}}\bigg)^{1/2},
\end{equation}
where we have normalized on the rate for production via the inverse decays of muons and taken the singlet scalar's mass to be a GeV. Crucially, values of the mixing angle required to explain the Belle II signal are above this critical angle and therefore we conclude that the scalar is brought into thermal equilibrium with the SM plasma. Moreover, since $\Gamma(h_D\to \chi\bar \chi)>\Gamma(h_D\to f\bar f)$ [cf. Fig.~\ref{fig:yD}], the dark fermions are also brought into equilibrium.

The freeze-out abundance of the fermions is then determined by their annihilation rate into SM particles. In this model, this occurs through $s$-channel $h_D$ exchange. Taking, e.g., $m_\chi=100~\rm MeV$, this annihilation is primarily into $e^+e^-$ pairs with a cross section
\begin{equation}
\begin{aligned}
\sigma v_\chi&\simeq \frac{y_D^2\sin^2\theta m_e^2}{4\pi v^2}\frac{m_\chi^2}{m_{h_D}^4}v_\chi^2
\\
&\simeq 4\times 10^{-47}~\frac{\rm cm^3}{\rm s}\bigg(\frac{y_D}{10^{-4}}\bigg)^2\bigg(\frac{\sin\theta}{10^{-3}}\bigg)^2\\
&\quad\quad\times\bigg(\frac{m_\chi}{100~\rm MeV}\bigg)^2\bigg(\frac{1~\rm GeV}{m_{h_D}}\bigg)^4.
\end{aligned}
\end{equation}
This ($p$-wave) annihilation cross section is rather small and leads to an extremely large relic abundance,
\begin{equation}
\begin{aligned}
\Omega_\chi h^2&\sim 10^{20}\bigg(\frac{10^{-4}}{y_D}\bigg)^2\bigg(\frac{10^{-3}}{\sin\theta}\bigg)^2\\
&\quad\quad\times\bigg(\frac{100~\rm MeV}{m_\chi}\bigg)^2\bigg(\frac{m_{h_D}}{1~\rm GeV}\bigg)^4,
\end{aligned}
\end{equation}
which overcloses the Universe. Therefore, we have to either introduce a new $\chi$ annihilation channel that does not involve $h_D$ or allow $\chi$ to decay.

\subsection{Modified cosmology and freeze-in}
Can we modify the cosmology simply to avoid this conclusion? One possibility is to avoid bringing the singlet scalar into equilibrium. This could be done, e.g., by reheating the Universe to a temperature $T<m_{h_D}$ or by arranging $h_D$ to be much heavier in the early Universe through thermal effects or a phase transition. In this case $\chi$s are produced at $T\sim m_\chi$ through freeze-in, $f\bar f\to\chi\bar\chi$. The abundance that is obtained in this case is proportional to the $\chi\bar \chi\to f\bar f$ cross section,
\begin{equation}
\begin{aligned}
\Omega_\chi h^2&\sim 2\times 10^{-3}\bigg(\frac{y_D}{10^{-4}}\bigg)^2\bigg(\frac{\sin\theta}{10^{-3}}\bigg)^2\\
&\quad\quad\times\bigg(\frac{m_\chi}{100~\rm MeV}\bigg)^4\bigg(\frac{1~\rm GeV}{m_{h_D}}\bigg)^4,
\end{aligned}
\end{equation}
where we have considered the production through $\mu^+\mu^-$ annihilation. Considering production through quarks would increase this but not enough for $\chi$ to make up the total observed dark matter density.

Note that it is possible to obtain the correct dark matter abundance for $\chi$ through freeze-in from the decays of $h_D$'s that have been brought into equilibrium. However, this requires $y_D\sim 10^{-12}$ \cite{Bauer:2017qwy,Bernal:2017kxu,Hall:2009bx} which, as we see in Fig.~\ref{fig:yD}, would make the invisible decay of $h_D$ extremely subleading, such that the Belle II signal could not be explained.

We find that the dark fermions can not obtain the correct relic abundance without either modifying the standard cosmology or introducing further interactions. Below, we will discuss a model that naturally includes new interactions so that the states into which $\hD$ decays are themselves unstable cosmologically.

\section{Simplest dark sector}\label{sec:DarkHiggs}

\begin{figure*}[tbh]
\includegraphics[width = 0.55\textwidth]{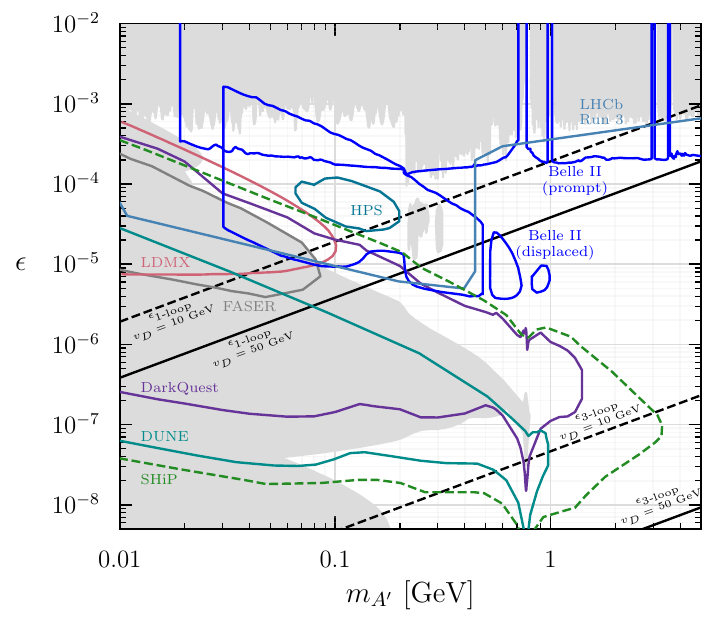}
\caption{Existing and future constraints on visibly decaying dark photons. The gray shaded region are existing constraints from \cite{Riordan:1987aw,Bjorken:1988as,Davier:1989wz,Bross:1989mp,Bjorken:2009mm,Blumlein:2011mv,APEX:2011dww,Andreas:2012mt,Blumlein:2013cua,BaBar:2014zli,Merkel:2014avp,Curtin:2014cca,NA482:2015wmo,BESIII:2017fwv,LHCb:2017trq,KLOE-2:2018kqf,NA64:2019auh,LHCb:2019vmc,CMS:2019buh,Tsai:2019buq,FASER:2023tle}. Future projections from Belle II \cite{Belle-II:2018jsg,Ferber:2022ewf}, DarkQuest \cite{Berlin:2018pwi}, DUNE \cite{Berryman:2019dme}, HPS \cite{Baltzell:2022rpd}, LDMX \cite{Berlin:2018bsc}, LHCb \cite{Craik:2022riw}, and SHiP \cite{SHiP:2020vbd} are shown by the various colored curves. The solid black (dashed) lines are the expected $1-$ and $3-$loop values of kinetic mixing for $v_D = 10 (50)$ GeV.} \label{fig:darkphotonbounds}
\label{fig:diagram}
\end{figure*}

A well-motivated model in which the dark sector states are naturally unstable is a dark $U(1)_D$ extension of the SM. In this model, the scalar $S$ is a component of a dark field $\Phi_S$ that has a $U(1)_D$ charge, and spontaneously breaks the gauge symmetry when it gets a VEV.  This generates a mass for the dark photon $A_D$. In this scenario, we choose the masses so that the dark Higgs decays to dark photons, which are unstable and decay to SM fermions. The dark photons can be long-lived enough such that they appear as missing energy at Belle II.\footnote{This model is the hidden Abelian Higgs model \cite{Wells:2008xg,Curtin:2014cca}, but since we are interested in processes that happen at energies much below the electroweak scale ($\sim M_W$) we consider dark photon mixing with the SM photon only.}

The relevant terms in the Lagrangian for this model are
\begin{align}\label{eq:LagDark}
\mathcal{L} &\supset -e\epsilon Q_f \bar{f} \gamma^\mu f A^{\prime}_\mu + |(\partial_\mu + i g_D A_{D\mu}) \Phi|^2  \nonumber \\
&+\mu_H (H^\dagger H) - \lambda_H (H^\dagger H)^2 \nonumber \\
& +\mu_S \Phi^2_S - \lambda_S \Phi_S^4 - \lambda_{SH} \Phi_S^2 (H^\dagger H),
\end{align}
where we have already made the minimal substitution $A_\mu \to A_\mu + \epsilon A_{D\mu}$ to diagonalize the kinetic terms for the SM photon $A_\mu$ and the dark photon $A_{D\mu}$. The dark Higgs fields can be written in the unitary gauge as
\begin{equation}
\Phi_S =\frac{1}{\sqrt{2}} (v_D + S)
\end{equation}
where $v_D$ is the VEV of dark Higgs.

The scalar $S$ and the SM Higgs $H$ mix as in Eq.~\eqref{eq:mixing} to form mass eigenstates $h$ and $h_D$, with the mixing angle now given by

\begin{equation}
\sin\theta \approx \frac{\lambda_{SH} v v_D}{m_h^2},
\end{equation}

and the masses of the physical states 
\begin{equation}
m_h^2 \approx 2 \lambda_H v^2,~m_{\hD}^2 \approx 2\lambda_S v_D^2.
\end{equation}

Stability of the scalar potential is guaranteed if (see App.~\ref{app:stability} for more details)

\begin{equation}\label{eq:stability}
4\lambda_H \lambda_S - \lambda_{SH}^2 >0, ~\text{and}~\lambda_H,\lambda_S > 0.
\end{equation}

In terms of the physical masses of the Higgs bosons and the mixing angle, the stability condition results in the requirement $m_{\hD}> m_h \sin\theta$, or 
\begin{equation}
m_{\hD} \gtrsim 0.7~{\rm GeV}\, \bigg(\frac{\sin\theta}{6\times 10^{-3}}\bigg),
\end{equation}
where we have normalized to the value of the mixing angle needed to obtain the central value of the NP contribution to $B^+ \to K^+ +{\rm inv.}$

After symmetry breaking, the dark photon acquires a mass $m_{A_D} = g_D v_D$, and the interactions of physical Higgs bosons to dark photons are 
\begin{equation}
\mathcal{L} \supset g_D^2 v_D A_\mu^\prime A^{\prime \mu} (-h\sin\theta  + h_D\cos\theta ).
\end{equation}

We can exchange the parameters of the Lagrangian for physical parameters i.e.

\begin{equation}
\{ \lambda_H, \lambda_S, \lambda_{SH}, g_D\} \to \{ m_h, m_S, \sin\theta, m_{A_D}\},
\end{equation}
which, together with $v,v_D$, are the set of parameters that we can vary. We fix $m_h = 125$~GeV and $v=246$~GeV, and we will consider different fixed values of $v_D$. Then, the remaining parameters to scan over are $m_{\hD}$ and $\sin\theta$. In this way, everything from the singlet Higgs portal scenario applies directly.

\subsection{Dark Photon Parameter Space}

Before discussing the effect of this model on $B^+ \to K^++{\rm inv.}$, we first consider the $\epsilon$ vs $m_{A_D}$ parameter space of the dark photon to determine when it is long-lived enough to appear as missing energy at Belle II. The dark photon couples to SM fermions via kinetic mixing with the SM photon, and is given by the first term $\mathcal{L} \supset -e\epsilon Q_f \bar{f} \gamma^\mu f A^{\prime}_\mu$ in Eq.~\eqref{eq:LagDark}. The decay rate of the dark photon to a pair of leptons when $m_{A_D} \gg m_\ell$ is 
\begin{equation}
\begin{aligned}
&\Gamma(A_D \to \ell^+ \ell^-) \simeq \frac{\epsilon^2 e^2}{12 \pi} m_{A_D}
\\
&=\frac{1}{3\times 10^{-5}~\rm s}\bigg(\frac{\epsilon}{10^{-8}}\bigg)^2\bigg(\frac{m_{A_D}}{100~\rm MeV}\bigg),
\end{aligned}
\end{equation}
while the decays to hadronic final states are given by
\begin{equation}
\Gamma(A_D \to \text{hadrons}) \simeq \Gamma(A_D \to \mu^+ \mu^-) R(\sqrt{s} = m_{A_D}),
\end{equation}
where $R \equiv \sigma(e^+ e^- \to \text{hardrons})/\sigma(e^+ e^- \to \mu^+ \mu^-)$ \cite{Workman:2022ynf}

We show existing and future bounds on visibly decaying dark photons in Fig.~\ref{fig:darkphotonbounds}, where the gray shaded regions depict existing constraints. We see that there are two regions of the $\epsilon-m_{A_D}$ plane for sub-GeV dark photons that are unconstrained by existing experimental searches:

\begin{itemize}
\item[1.] $m_{A_D} \gtrsim $ 100 MeV with $10^{-8} \lesssim \epsilon \lesssim 10^{-7}$
\item[2.] $m_{A_D} \gtrsim$ 50 MeV with $10^{-6} \lesssim \epsilon  \lesssim 10^{-4}$.
\end{itemize}

At Belle II, dark photons are produced in association with a SM photon $e^+ e^- \to \gamma A_D$, and can decay to a pair of charged particles if there are no lighter dark sector states for the dark photon to decay into. In the first region, the dark photon has a decay length that is much larger than the Belle II detector and escapes as missing energy. Limits on invisibly decaying dark photons with $m_{A_D} \gtrsim 100$ MeV from NA64 \cite{Andreev:2021fzd} and BaBar \cite{BaBar:2017tiz} constrain the kinetic mixing to be $\epsilon \lesssim 10^{-3}$, and Belle II is projected to improve this constraint to $\epsilon \lesssim 3\times10^{-4}$ \cite{Belle-II:2018jsg,Graham:2021ggy}.

In the second region, it is possible that the dark photon appears visibly as a displaced vertex at Belle II. A recent study on the Belle II sensitivity to long-lived dark photons was presented in \cite{Ferber:2022ewf}. In this work it was shown that decay lengths up to about $60~\rm cm$, corresponding to $\epsilon\gtrsim 10^{-5}$, could be probed by reconstructing displaced vertices. We show the projected sensitivity in Fig.~\ref{fig:darkphotonbounds} with the blue curves labeled ``Belle II (displaced)''. Longer decay lengths up to a couple of meters could potentially be probed in the outer layers of the Belle II detector. 

For our purposes, the first region above would safely result in the dark photons appearing as missing energy in the $b\to s$ decays at Belle II. In the second region, one could ensure that the dark photons appear as missing energy by introducing a new state for them to decay into. Interestingly, this state could be the dark matter and can be searched for at experiments such as NA62 and the upcoming LDMX~\cite{LDMX:2018cma,Graham:2021ggy}.

One can also ask what values of kinetic mixing should be expected given the requirement on the level of Higgs mixing to explain the Belle II signal. It is reasonable to assume that there is matter charged under both $U(1)_D$ and electromagnetism and that in the ultraviolet the mixings $\lambda_{HS}$ and $\epsilon$ vanish. Nonzero values for these mixings arise after integrating out this heavy matter. This leads to the Higgs portal interaction at one-loop,
\begin{equation}
\lambda_{SH}\sim\frac{\lambda^2\lambda_D^2}{16\pi^2},
\end{equation}
where $\lambda$ and $\lambda_D$ are Yukawa couplings. In such a case it is also reasonable to assume that the kinetic mixing arises at one loop from integrating out this matter, leading to

\begin{equation}
\begin{aligned}
\epsilon_{1-\rm loop}&\sim\frac{eg_D}{16\pi^2}
\\
&\sim 4\times10^{-6}\bigg(\frac{m_{A_D}}{100~\rm MeV}\bigg)\bigg(\frac{50~\rm GeV}{v_D}\bigg).
\end{aligned}
\end{equation}
It is also possible to write models involving neutrino-portal-type interactions that generate kinetic mixing only at 3-loops through the Higgs mixing term~\cite{Gherghetta:2019coi},
\begin{equation}
\begin{aligned}
&\epsilon_{3-\rm loop}\sim\frac{eg_D}{16\pi^2}\times\frac{\lambda^2\lambda_D^2}{\left(16\pi^2\right)^2}\simeq\frac{eg_D}{16\pi^2}\times\frac{m_h^2\sin\theta}{16\pi^2v\, v_D}
\\
&\sim 10^{-10}\bigg(\frac{m_{A_D}}{100~\rm MeV}\bigg)\bigg(\frac{50~\rm GeV}{v_D}\bigg)^2\bigg(\frac{\sin\theta}{10^{-3}}\bigg),
\end{aligned}
\end{equation}
where we have re-expressed the Yukawa couplings in terms of the Higgs mixing angle. We show the expected 1- and 3-loop values of kinetic mixing in Fig.~\ref{fig:darkphotonbounds} for $v_D=10$,~$50~\rm GeV$ and $\sin\theta=6\times10^{-3}$. We see that these expected contributions can accommodate both of the benchmark regions of mass and mixing mentioned above.

Future experimental searches will shed light on these benchmark regions as well. In Fig.~\ref{fig:darkphotonbounds}, we show the projected sensitivity of other experiments by the colored curves.  DUNE \cite{Berryman:2019dme} (cyan curve) and SHiP \cite{SHiP:2020vbd} (dashed green curve) can probe visibly decaying dark photons with kinetic mixing of a few $\times 10^{-8}$, due to their high intensity and long decay volumes. SHiP and DarkQuest \cite{Berlin:2018pwi} (purple curve) have sensitivity to visible dark photons with shorter lifetimes, with kinetic mixing of $\sim 10^{-5}$. 
 
 For the remainder of this paper, we fix $m_{A_D} = 100~\rm MeV$ and assume that the kinetic mixing is in the two windows discussed above, such that the dark Higgs decays to dark photons 100\% of the time (when kinematically allowed), and  the dark photons appear as missing energy at Belle II.

\subsection{Dark Higgs contrution to $B^+ \to K^++{\rm inv.}$}

\begin{figure*}[t]
\includegraphics[width = 0.48\textwidth]{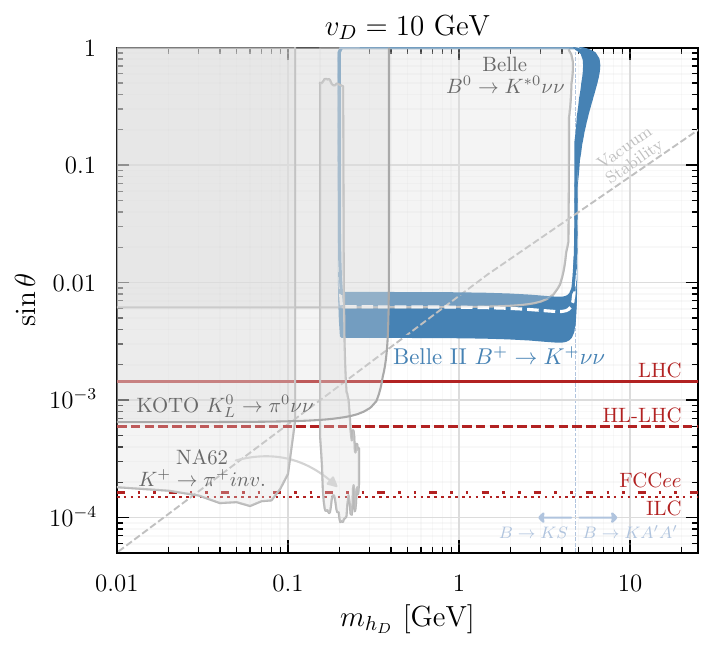}~
\includegraphics[width = 0.48\textwidth]{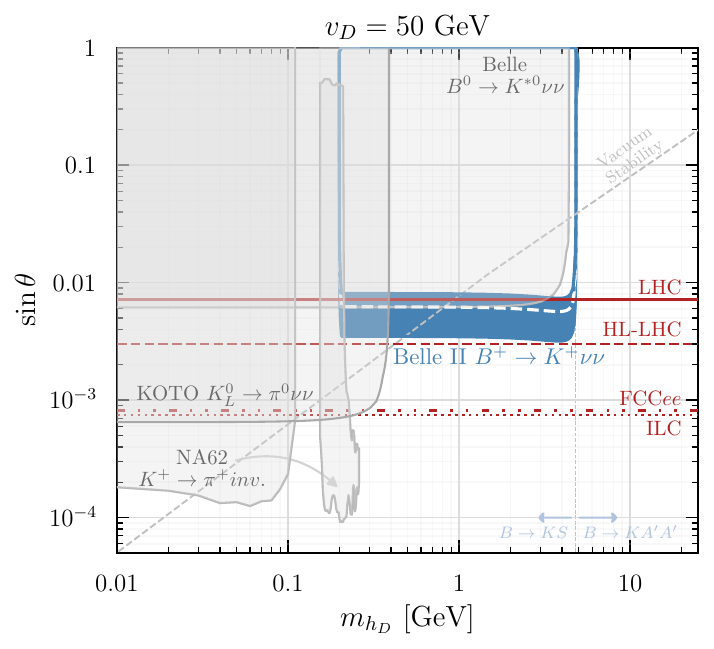}~
\caption{Same as Fig.~\ref{fig:bounds} except for the Dark Abelian Higgs Model with $v_D = 10$ GeV (left) and $v_D = 50$ (right). In both plots we fix the dark photon mass to $m_{A_D} = 100$ MeV. The dashed gray line depicts the stability condition of Eq.~\eqref{eq:stability}.}
\label{fig:bounds_DP}
\end{figure*}

When the dark Higgs is lighter than the $B$ mesons, it is produced on-shell in two-body decays and Eqs.~\eqref{eq:BpltoKpl}-\eqref{eq:KtoPi} can be used to determine the preferred regions of parameter space for the Belle II results and constraints from other $B,K$ decays.  When $m_{\hD} > m_B$, $B$ meson decays to a pair of dark photons are mediated by the SM Higgs and the dark Higgs. The differential decay rates are given by
%
\begin{widetext}
\begin{align}
\frac{d\Gamma}{dq^2} (B^+ \to K^+ A_D A_D)&= \frac{ |f^{BK}_0(q^2)|^2 }{8192 \pi^3 m_{B^+}^3}\bigg( \frac{m_{B^+}^2 - m_{K^+}^2}{m_b - m_s}\bigg)^2  \bigg|\frac{c_{bs}\sin\theta\cos\theta}{q^2 - m^2_{\hD} + i m_{\hD} \Gamma_{\hD}} - \frac{c_{bs}\sin\theta\cos\theta}{q^2 - m^2_h}\bigg|^2\nonumber\\
&\hspace{3cm}\times\frac{q^4 - 4m_{A_D}^2 q^2 + 12 m_{A_D}^4}{v_D^2}  \bigg(1 - \frac{4m_{A_D}^2}{q^2}\bigg)^{1/2} \lambda^{1/2}(q^2,m_{B^+}^2, m_{K^+}^2),\\
\frac{d\Gamma}{dq^2} (B^0 \to K^{\ast 0} A_D A_D)&= \frac{ |A^{BK}_0(q^2)|^2 }{8192 \pi^3 m_{B^0}^3}\frac{1}{(m_b + m_s)^2}  \bigg|\frac{c_{bs}\sin\theta\cos\theta}{q^2 - m^2_{\hD} + i m_{\hD} \Gamma_{\hD}} - \frac{c_{bs}\sin\theta\cos\theta}{q^2 - m^2_h}\bigg|^2\nonumber\\
&\hspace{3cm}\times\frac{q^4 - 4m_{A_D}^2 q^2 + 12 m_{A_D}^4}{v_D^2}  \bigg(1 - \frac{4m_{A_D}^2}{q^2}\bigg)^{1/2} \lambda^{3/2}(q^2,m_{B^0}^2, m_{K^{\ast 0}}^2),
\end{align}
\end{widetext}
where, again, $\Gamma_{\hD}$ is the total decay rate of the the dark Higgs and is dominated by decays to dark photons when $v_D \lesssim (m_\mu m_h/v\sin\theta)^{-1}$. The decay rate for $h_D \to A_D A_D$ is
\begin{equation}   
\begin{aligned}
\Gamma(h_D \to A_D A_D) &= \frac{m_{\hD}^4 - 4m_{\hD}^2m_{A_D}^2 - 12 m_{A_D}^4}{32 \pi m_{\hD}v_D^2} \\
&\hspace{2cm}\times \sqrt{1-\frac{4m_{A_D}^2}{m_{\hD}^2}}.
\end{aligned}
\end{equation}
For mixing angles that can address the Belle II results for $B^+ \to K^+ \nu\nu$, we find that $v_D \lesssim 3$ TeV and the dark Higgs decays invisibly.

In Fig.~\ref{fig:bounds_DP} we show the region of the $\sin\theta -m_{\hD}$ plane that can explain the Belle II excess by the blue shaded region, for $v_D = 10$ GeV (left plot) and $v_D = 50$ GeV (right plot) and a fixed dark photon mass $m_{A_D} = 100$ MeV. Similar to the singlet Higgs scenario of Sec.~\ref{sec:HiggsPortal}, the Belle II result can be explained when the mixing angle $\sin\theta \simeq 6\times 10^{-3}$ and a dark Higgs mass $m_{\hD} \lesssim m_B$. However, stability of the scalar potential required by Eq.~\eqref{eq:stability} requires that $m_{\hD} \gtrsim 500$ MeV. We show this by the dashed gray line in both plots of Fig.~\ref{fig:bounds_DP}.

In addition, bounds from $K^+ \to \pi^+ + {\rm inv.}$~\cite{NA62:2021zjw}, $K_L^0 \to \pi^0 \nu\nu$~\cite{KOTO:2020prk}, and $B^0 \to K^{\ast 0} \nu\nu$~\cite{Belle:2017oht} are depicted bye the gray shaded regions. Similar to the singlet Higgs portal scenario, we observe that searches for invisible kaon decays exclude $m_{\hD}\lesssim 400$ MeV, while bounds from $B^0 \to K^{\ast 0} \nu\nu$ exclude some of the parameter space that can explain the Belle II excess.

\subsection{Higgs Invisible Decays}
Compared to the singlet Higgs portal model in Sec.~\ref{sec:HiggsPortal}, invisible decays of the SM Higgs boson has contributions from  both dark photons $h \to A_D A_D$ and dark Higgs bosons $h \to h_D h_D$. The decay widths for these are
\begin{align}
\Gamma(h \to A_D A_D) &= \frac{\sin^2\theta}{32 \pi m_h}\frac{m_h^4 - 4m_h^2m_{A_D}^2 - 12 m_{A_D}^4}{v_D^2} \nonumber\\
&\hspace{3cm}\times \sqrt{1-\frac{4m_{A_D}^2}{m_h^2}},\\
\Gamma(h \to h_D h_D) &= \frac{\sin^2\theta}{64 \pi}\frac{m_h^3}{v_D^2}\sqrt{1-\frac{4m_{h_D}^2}{m_h^2}}.
\end{align}

Bounds on SM Higgs invisible BR from LHC (HL-LHC) in Tab.~\ref{tab:Hinv} require
\begin{equation}
\frac{\sin\theta}{v_D} \lesssim 1.3\times 10^{-4}~(5.9 \times 10^{-5})~\text{GeV}^{-1},
\end{equation}
and are depicted in Fig.~\ref{fig:bounds_DP} by the solid and dashed red lines, respectively. For $v_D = 10$ GeV we see that current LHC bounds on Higgs invisible decays exclude all of the parameter space favored by the Belle II excess. By increasing the dark Higgs VEV to $v_D = 50$ GeV, these bounds weaken and we see that the current LHC bounds are just starting to exclude the blue region in the right plot of Fig.~\ref{fig:bounds_DP}. In the future, HL-LHC will probe all of the parameter space for $v_D$ = 50 GeV. 

\section{Conclusion}\label{sec:conclusion}
We have studied the compatibility of the recent Belle II evidence for $B\to K+{\rm inv.}$ with new physics that couples to the SM through the Higgs portal. We have shown that, unless the Higgs sector is extended, limits on the invisible branching of the Higgs boson force the new states to be below the $B$ mass. If this excess holds up, kinematic information using the large expected Belle II dataset in the future might be able to shed light on the mass of the light state produced in $B\to K$ decays.

Since the signal we study involves missing energy, it is natural to ask whether it can be the dark matter required by cosmological and astrophysical observation. We find that generically, the new states must be coupled strongly enough to be brought into equilibrium with the SM plasma in the early Universe, but not strongly enough to deplete their density to acceptable levels if they are cosmologically stable.

Motivated by this, we studied a simple, well-motivated dark sector consisted of a $U(1)_D$ gauge symmetry that is spontaneously broken by the VEV of a scalar field. This scalar can mix with the SM Higgs boson with the required strength to be produced in $b\to s$ transitions and subsequently decay into the $U(1)_D$ gauge bosons. These gauge bosons, dark photons, can either decay back into SM states with a lifetime long enough to appear as missing energy at Belle II or into lighter dark sector states that could comprise the dark matter. The mass and kinetic mixing of the dark photons required by the Belle II excess falls into allowed regions of parameter space that are also well-motivated by top-down considerations.

A number of upcoming experiments and measurements are well poised to probe this explanation of the Belle II signal. In particular, the HL-LHC and proposed Higgs factories will improve the limits on the invisible Higgs width by an order of magnitude, directly testing the Higgs portal interpretation of the $B\to K+{\rm inv.}$ signal. Additionally, in the case that the Higgs-mixed scalar decays into dark photons, searches for long-lived particles at DUNE, SHiP, DarkQuest, and Belle II will target favored regions of $\epsilon$ and $m_{A^\prime}$. Furthermore, missing-momentum searches such as NA62 and LDMX could probe the possibility that the dark photon decays into dark matter.

Note, while this work was being completed, a related study that considered light new vector and axion-like particles as explanations for the Belle II results appeared \cite{Altmannshofer:2023hkn}. 

\acknowledgements
DT thanks the Carleton University Particle Theory Group for helpful discussions in the very early stages of this work. DM and DT are supported by Discovery Grants from the Natural Sciences and Engineering Research Council of Canada (NSERC). TRIUMF receives federal funding via a contribution agreement with the National Research Council (NRC) of Canada.

\appendix

\section{Form Factors in $B,K$ meson decays}\label{app:formfactors}

The hadronic matrix elements for $B \to K, K^\ast$ decays, where $K$ and $K^\ast$ denote a pseudoscalar and vector kaons, respectively, are given by \cite{Gubernari:2018wyi}
%
\begin{align}
\langle K(k)| \bar{q} b| B(p)\rangle &= \frac{m_B^2 - m_K^2}{m_b - m_s} f^{BK}_0(q^2) \\
\langle K(k)| \bar{q} \gamma_5 b| B(p)\rangle &= 0 \\[10pt]
\langle K^\ast(k)| \bar{q} b| B(p)\rangle &=0 \\
\langle K^\ast(k)| \bar{q} \gamma_5 b| B(p)\rangle &= -i \epsilon^\ast_{K^\ast,\nu}q^\nu \frac{2 m_{K^\ast}}{m_b+ m_s} A^{BK}_0 (q^2),
\end{align}
%
where $q^2 = (p-k)^2$, and $f_0, A_0$ are $q^2$-dependent form factors. Similarly, for the decays of pseudoscalar kaons to pions $K\to \pi$ decay we have

\begin{equation}
\langle \pi(k)| \bar{q} b| K(p)\rangle = \frac{m_K^2 - m_\pi^2}{m_s - m_d} f^{K\pi}_0(q^2).
\end{equation}

Note, the $K \to \pi$ form factor $f_0^{K \pi}$ is close to unity for the $0 < q^2 \lesssim (m_K - m_\pi)^2$ \cite{Boyle:2010bh,Carrasco:2016kpy}.

The form factors for $B$ meson decays can be written in the Bharucha-Straub-Zwicky (BSZ) parameterization as \cite{Bharucha:2015bzk}%
\begin{equation}
F_i(q^2) =  \frac{1}{1-q^2/m_R^2} \sum_{k = 0,1,2} a_k \big[ z(q^2) - z(0) \big]^k,
\end{equation}
where $m_R$ is a resonance mass, and

\begin{equation}
z(t) = \frac{ \sqrt{t_+ - t} - \sqrt{t_+ - t_0} }{\sqrt{t_+ - t} + \sqrt{t_+ - t_0} },
\end{equation}

with $t_0 = t_+(1 - \sqrt{1 - t_-/t_+})$ and $t_\pm = (m_B \pm m_{P,V})^2$. The inputs parameters $a_k$ and $m_R$ are given in Tab.~\ref{tab:FFs} \cite{Athron:2023hmz}.

\begin{table}[tbh]
\begin{ruledtabular}
\begin{tabular}{l c c c c}
 & $m_R$ [GeV] & $a_0$ & $a_1$ & $a_2$\\
\hline
$B \to K$ & 5.630 & 0.332& 0.335& 3.72$\times10^{-3}$\\
$B \to K^\ast$ & 5.336 & 0.342 &-1.147 & 2.372\\
\end{tabular}
\end{ruledtabular}
\caption{\label{tab:FFs} Parameters for $B\to P,V$ form factors in the BSZ parmaterization.}
\end{table}


\section{Minimization and Stability of the Scalar Potential of the Dark Abelian Higgs Model}\label{app:stability}

The scalar potential of the Dark Abelian Higgs Model is 
\begin{align}\label{eq:Vhiggs}
V(H,\Phi_S) &= -\mu_H (H^\dagger H) + \lambda_H (H^\dagger H)^2 \nonumber \\
& -\mu_S \Phi^2_S + \lambda_S \Phi_S^4 + \lambda_{SH} \Phi_S^2 (H^\dagger H),
\end{align}

and minimum of the scalar potential is found by taking first derivatives with respect to the scalar fields. This is equivalent to taking derivatives with respect to the VEVs and setting all field values to zero \cite{Pruna:2013bma,Robens:2015gla}, and we obtain
\begin{align}
\frac{\partial V}{\partial v} &= v(-\mu_H^2 + \lambda_H v^2 + \frac{1}{2} \lambda_{SH} v_D^2) = 0\\
\frac{\partial V}{\partial v_D} & =v_D(-\mu_S^2 + \lambda_s v_D^2 + \frac{1}{2} \lambda_{SH} v^2) = 0,
\end{align}
which we can use to rewrite the mass paramaters $\mu^2_{S,H}$.  

Alternatively, we can use these equations to get expressions for the VEVs:

\begin{equation}
v = \frac{\lambda_S \mu_H^2 - \frac{1}{2}\lambda_{SH} \mu_S^2}{\lambda_H \lambda_S -\frac{1}{4} \lambda_{SH}^2},~v_D =\frac{\lambda_H \mu_S^2 - \frac{1}{2}\lambda_{SH} \mu_H^2}{\lambda_H \lambda_S -\frac{1}{4} \lambda_{SH}^2},
\end{equation}
and we can that to have positive-definite VEVs requires
\begin{align}
\lambda_S \mu_H^2 - \frac{1}{2}\lambda_{SH} \mu_S^2 &> 0,\\
\lambda_H \mu_S^2 - \frac{1}{2}\lambda_{SH} \mu_H^2 &> 0
\end{align}

Stability of the scalar potential is found by taking the determinant of the Hessian matrix, and we have
\begin{align}
\frac{\partial^2 V}{\partial v^2}\frac{\partial^2 V}{\partial v_D^2}& -\bigg(\frac{\partial^2 V}{\partial v \partial v_D}\bigg)^2 \nonumber \\
&= v^2 v_D^2 (4\lambda_H \lambda_S - \lambda_{SH}^2) > 0.
\end{align}
Since $v,v_D \neq 0$, stability of the scalar potential is fulfilled if
\begin{equation}
4\lambda_H \lambda_S - \lambda_{SH}^2 >0, ~\text{and}~\lambda_H,\lambda_S > 0
\end{equation}



\bibliography{references}

\end{document}